\documentclass[review]{elsarticle}

\usepackage{lineno,hyperref}
\modulolinenumbers[5]

\journal{Cryogenics}

\begin{document}

\begin{frontmatter}

\title{A 10 mK hermetic cell for eliminating parasitic heating in cryogen-free dilution refrigerators}
%\tnotetext[mytitlenote]{Fully documented templates are available in the elsarticle package on \href{http://www.ctan.org/tex-archive/macros/latex/contrib/elsarticle}{CTAN}.}

%% or include affiliations in footnotes:
\author[Neel,Charles]{David Schmoranzer}
\author[Neel]{Sumit Kumar}
\author[Neel]{S\'{e}bastien Triqueneaux}
\author[NRL]{Xiao Liu}
\author[NRL]{Thomas Metcalf}
\author[NRL]{Glenn Jernigan}
\author[Neel]{Eddy Collin}
\author[Neel]{Andrew Fefferman\corref{mycorrespondingauthor}}
\cortext[mycorrespondingauthor]{Corresponding author}
\ead{andrew.fefferman@neel.cnrs.fr}

\address[Neel]{Univ. Grenoble Alpes, CNRS, Grenoble INP, Institut N\'{e}el, 25 rue des Martyrs, 38042 Grenoble, France}
\address[Charles]{Charles University, Prague, Czech Republic}
\address[NRL]{Naval Research Laboratory, Washington, D.C., USA}

\begin{abstract}
Cryogen-free dilution refrigerators generally simplify low temperature research but some types of samples, including superconducting qubits and other nanoelectronic devices, are affected by environmental heat sources such as stray photons or residual helium. We present the design and performance of a hermetic cell installed on the mixing chamber plate of a cryogen-free dilution refrigerator. The performance was quantified by measuring the dependence of the resonance frequency of a mechanical oscillator installed inside the cell on the mixing chamber temperature down to 10 mK. We found the expected logarithmic temperature dependence of the resonance frequency down to the lowest temperatures, demonstrating that the efficiency of the hermetic shield is significantly better than that of a simpler shield with no visible gaps.
\end{abstract}


\end{frontmatter}

%\linenumbers

\section{Introduction}

In many cases, cryogen-free (dry) dilution refrigerators greatly increase the efficiency of low temperature research. The elimination of the helium bath allows for much more experimental space than in conventional (wet) dilution refrigerators  as well as completely automated operation. However, in some cases, special measures are required to isolate the experiment from vibrations produced by the cryocooler of the dry fridge \cite{Pelliccione13,Kalra16,Schmoranzer19c}. Furthermore, Schmoranzer \emph{et al}. previously reported that sample heating due to environmental sources is more significant in dry fridges than in wet fridges \cite{Schmoranzer19a}. Vibrations were ruled out as the cause of heating. It was speculated that the excess heating was related to the absence of a hermetic shield at 4 kelvin (IVC) in the dry fridge, which leads to additional stray photons or residual helium gas at the mixing chamber plate.

Several previous works reported deleterious effects of stray infrared light in the sample space. In experiments involving superconducting cavities and qubits, stray light is a significant source of non-equilibrium quasiparticles. For example, Barends \emph{et al}. reported an increase in the quality factor of aluminum microwave cavities from 10$^5$ to 10$^6$ upon improving the shielding of the sample from ``typical" to multistage \cite{Barends11}. These measurements were made in an electronic adiabatic demagnetization refrigerator. The optimal shielding configuration was a box-in-a-box design with a black coating: While a substantial decrease in resonator $Q$ was observed when warming the 4 K stage in the case of typical shielding, no dependence on 4 K stage temperature was observed in the case of optimal shielding. The latter configuration also produced a significant improvement in the relaxation rate of a phase qubit. Similarly, C\'{o}rcoles \emph{et al.} reported an increase in the relaxation time of a superconducting qubit after embedding its package in ECCOSORB absorptive epoxy \cite{Corcoles11}. In this case, the sample was attached to the mixing chamber plate of a dry dilution refrigerator. These effects are not limited to qubit measurements. Bradley \emph{et al.} cooled a Coulomb blockade thermometer (CBT) nanoelectronic device below 5 mK via on-chip magnetic cooling in a dry dilution refrigerator. In their detailed thermal model of the cooling process, it was speculated that a parasitic heat leak into the CBT resulted from non-equilibrium photons in the environment \cite{Bradley17}. Here we present the design of a hermetic cell connected to the mixing chamber plate of a dry dilution refrigerator and quantify its effectiveness in improving the thermalization of our sample, which is a mechanical resonator. We observe a significant improvement relative to a simpler shield described in \cite{Schmoranzer19a} that was not vacuum tight.

\section{Cell Design and Construction}
\begin{figure}
  \includegraphics[width=\textwidth]{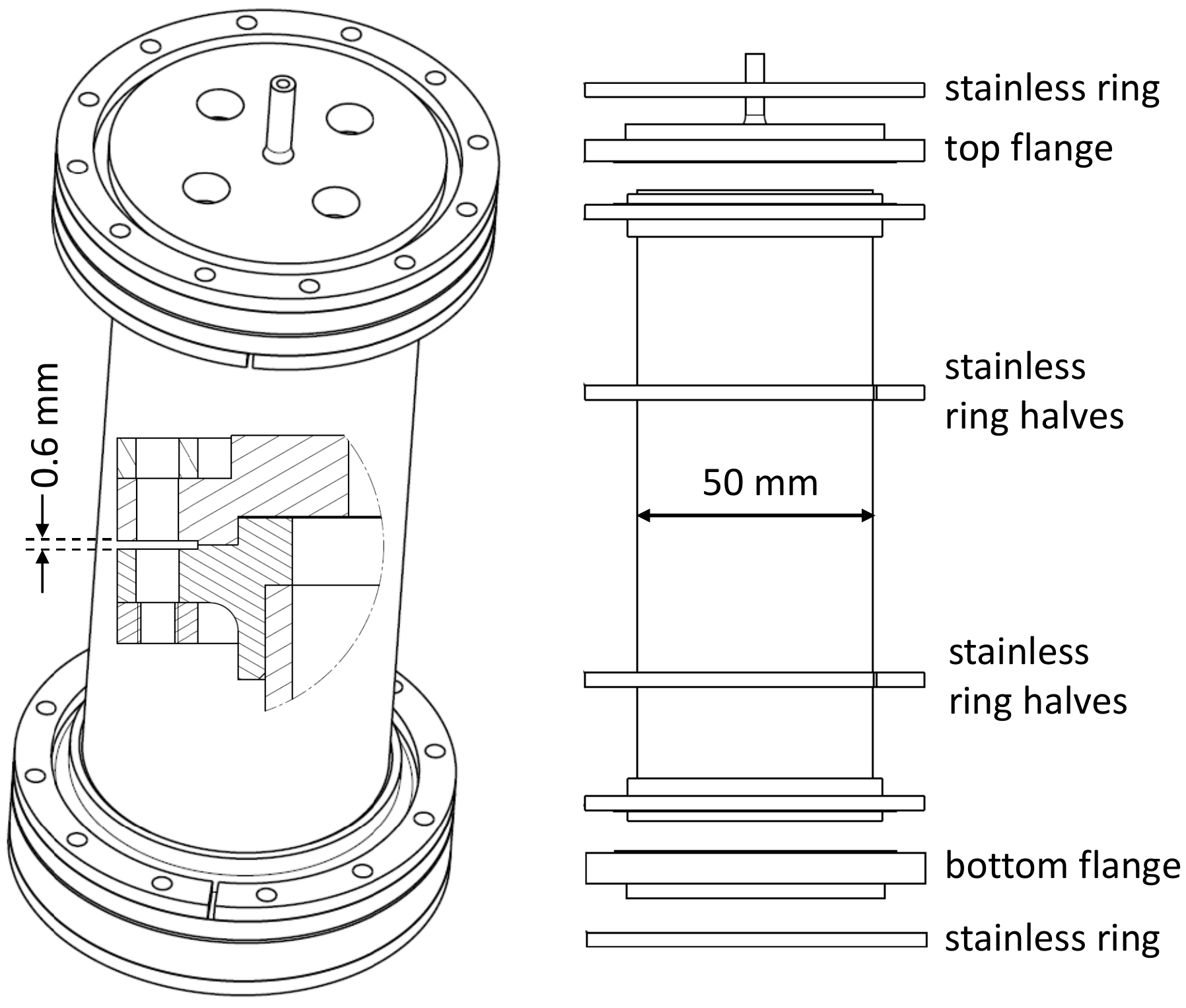}
\caption{Assembly drawing of the hermetic cell. The inset shows a cross-sectional detail of one of the indium-sealed flanges. The indicated 0.6 mm gap prevents interference of the flanges due to warping when compressing the indium ring.}
\label{cell}
\end{figure}

Our hermetic cell consists of a copper tube with indium-sealed flanges at each end (Fig. \ref{cell}). Ordinary copper was used for the tube and CuC2 (i.e. 3N5 purity Cu) was used for the flanges. An inner flange was welded to each end of the tube. Mating flanges were attached using 12 M3 stainless steel screws at each end. Stainless steel rings were used when closing the cell in order to minimize warping of the copper flanges (Fig. \ref{cell}). We would recommend using even thicker flanges to further suppress warping at the outer part of the flange.

\begin{figure}
  \includegraphics[width=7cm]{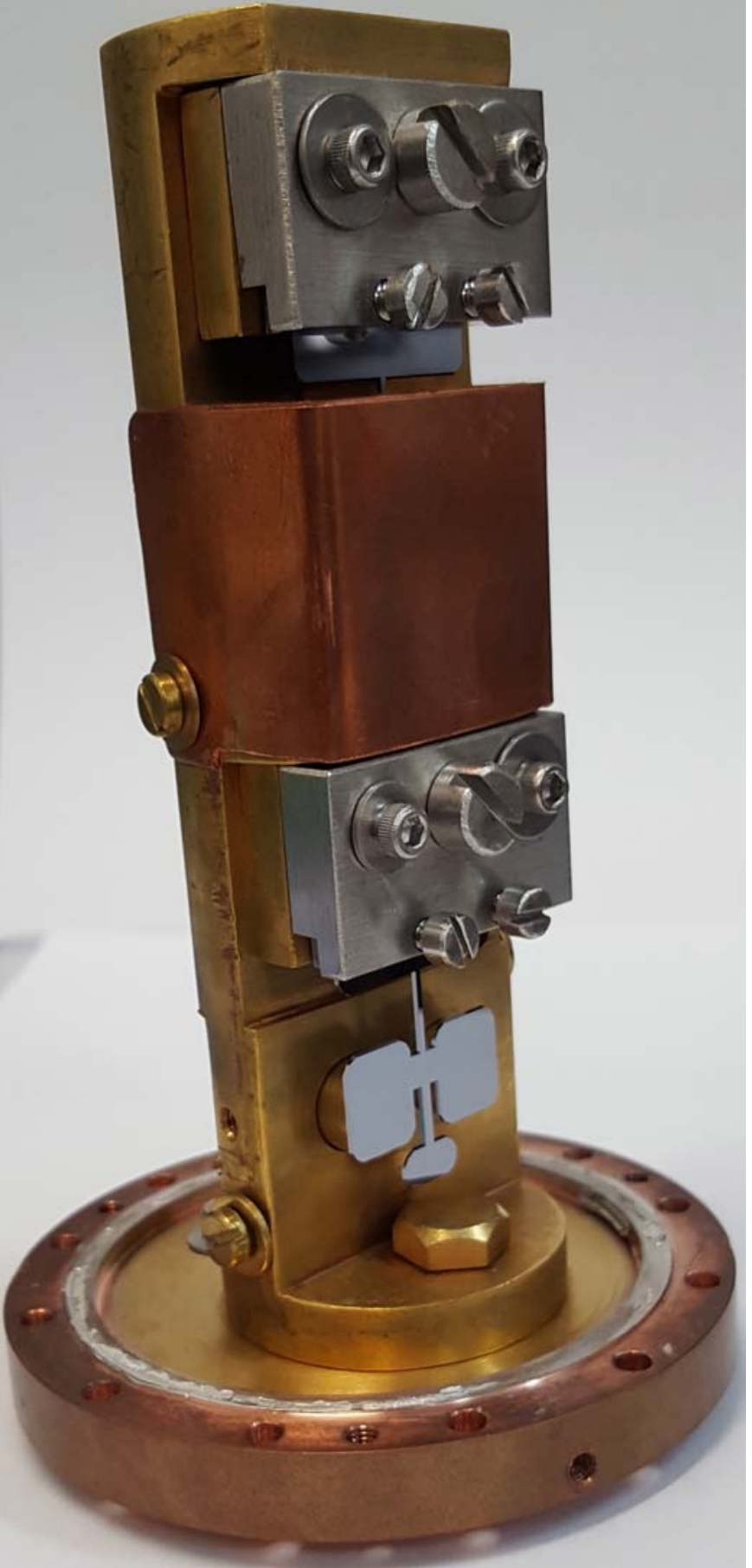}
\caption{The gold plated copper support structure of the mechanical resonator samples is screwed to the gold plated top surface of the bottom flange of the hermetic cell. The copper sheet in front of the upper mechanical resonator protects this fragile device from breakage during assembly of the cell. Another sheet that protects the lower mechanical resonator has been removed for the photo.}
\label{bot_flange}
\end{figure}

The bottom of the bottom flange has five M3 threaded holes used to connect the cell to the mixing chamber plate of our Bluefors LD400 cryogen-free dilution refrigerator. The opposite side of this flange, which is the vacuum side, has threaded holes used to attach the copper support structure of our mechanical oscillator. (Fig. \ref{bot_flange}). Both sides of the flange are gold plated to ensure a good thermal contact to the sample.

The top flange has four holes for coaxial feedthroughs and a central spout in which a copper pumping tube was brazed. The coaxial feedthroughs were double ended hermetic SMA bulkhead adaptors manufactured by Huber and Suhner (34\_SMA-50-0-3/111\_N). They were adapted for low temperature use by replacing the elastomer gasket with an indium ring. The pumping line was a 50 cm long copper capillary tube with 2 mm outer diameter and 0.8 mm inner diameter. After closing the cell, it was connected to a helium leak detector and pumped for two days with a turbomolecular pump. No leak was detected when spraying the indium joints and the SMA feedthroughs with helium. The hermetic cell was then sealed by pinching the pumping line between two cylindrical surfaces using a manual vise-grip tool with a lever arm of approximately 15 cm. The pumping line was pinched at three nearby locations at the end of tube, with the third pinch being forceful enough to sever the tube. As a precaution, some PbSn solder was melted onto the end of the severed pumping line. The cell has been closed and thermally cycled between ambient and mK temperatures several times and remains leak-tight.

\section{Cell Performance}
The performance of the cell was quantified using a mechanical oscillator sample called a double paddle oscillator (DPO). In particular, we measured the temperature dependence of the resonance frequency of a mechanical mode of the DPO near 5 kHz down to a mixing chamber plate temperature of 10 mK. The DPO consisted of a 300 $\mu$m thick crystalline silicon substrate with a 300 nm layer of amorphous silicon deposited on top. The lateral dimensions of the oscillator were on the order of 1 cm. The inverse quality factor $1/Q$ and the temperature dependence of the resonance frequency of the resonator is dominated by the contribution of the thin amorphous film. Even with the amorphous layer, $1/Q$ remains low, ranging from $4\times 10^{-8}$ at 10 mK to $5\times 10^{-7}$ at 1 K. A detailed description of the oscillator and the measurement technique is given in \cite{Schmoranzer19a}.

\begin{figure}
% Use the relevant command to insert your figure file.
% For example, with the graphicx package use
  \includegraphics[width=\textwidth]{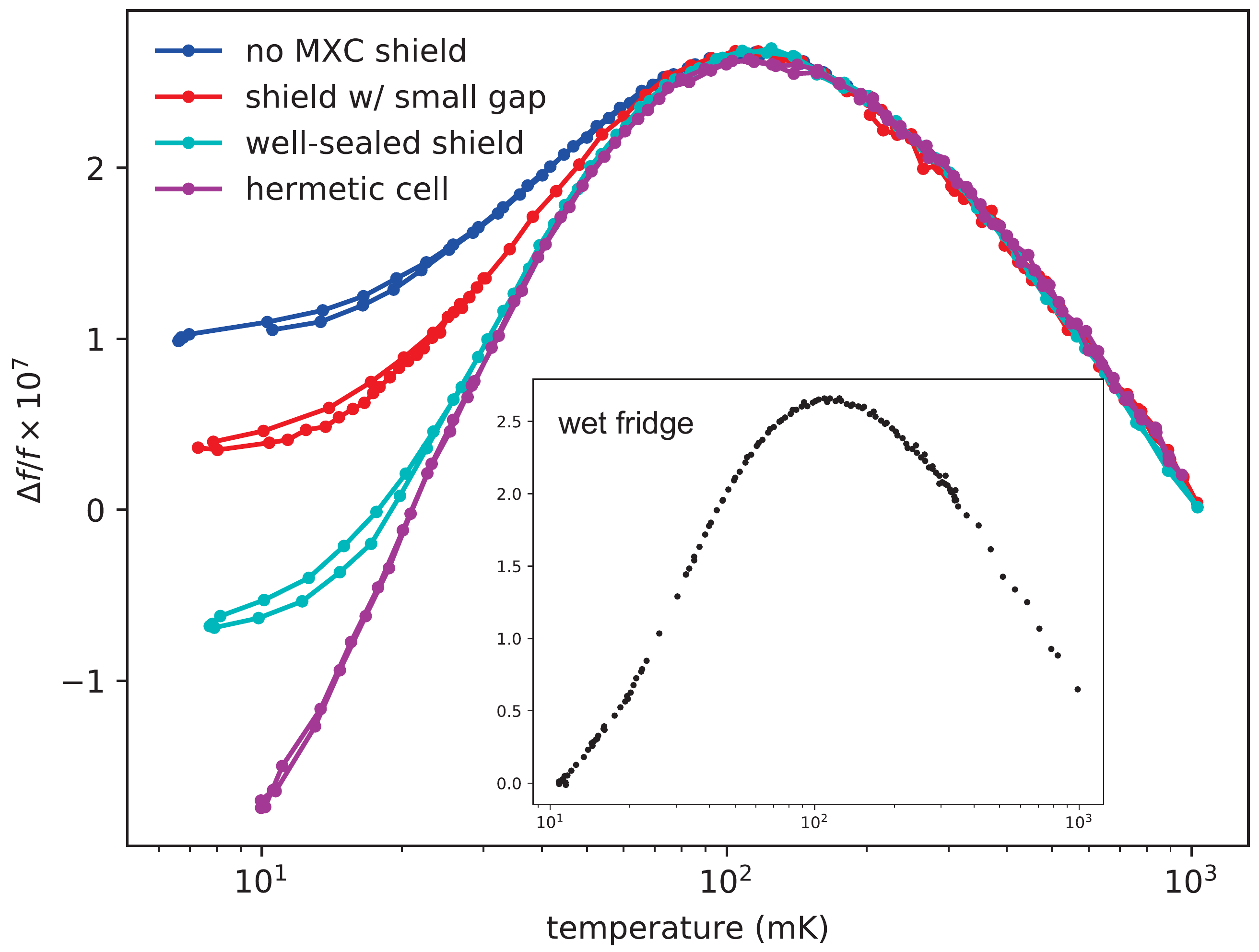}
% figure caption is below the figure
\caption{Relative frequency shift of one mode of a single crystal Si mechanical resonator with a 300 nm thick amorphous Si film deposited at room temperature. The labels correspond to different shield configurations (see text). Warming and cooling curves are shown for each shield configuration. For the measurements with the hermetic cell the temperature was varied slowly enough to eliminate the hysteresis visible in the other measurements. The inset shows the relative frequency shift of another mechanical resonator of the same type measured in a wet fridge.}
\label{fig:3}       % Give a unique label
\end{figure}

Figure \ref{fig:3} shows the temperature dependence of the resonance frequency
of the present sample (called DPO8) obtained from ringdown measurements like those shown in Fig.
 1 of \cite{Fefferman16}. Measurements of DPO8 in four configurations are shown in Fig. \ref{fig:3}: (1) no mixing chamber shield surrounding the DPO, (2) a simple mixing chamber shield with a $\approx$ 1 cm gap but no line of sight from the still shield to the DPO, (3) with all visible gaps in the simple shield closed with copper tape, and (4) the hermetic shield described above. A vertical offset was added to each of the DPO measurements so that the high temperature data overlap. The inset shows the temperature dependence of the resonance frequency of a very similar sample measured in a wet fridge. This sample was another DPO (called DPO1), also with a 300 nm layer of amorphous Si deposited at room temperature.

The temperature dependence of the resonance frequency of DPO8 below 50 mK strengthened as the mixing chamber shield was improved, demonstrating that the thermalization of the sample was improving. In the case of the hermetic cell, the logarithmic temperature dependence of the resonance frequency predicted by the standard tunneling model of amorphous solids \cite{Phillips87} is observed, indicating excellent sample thermalization down to 10 mK. The hermetic cell yields better thermalization of the sample than was achieved in the wet fridge, which had a hermetic seal at 4 K, a radiation shield attached to the still plate and no additional radiation shields inside the still shield.

An upper limit on the magnitude of heat required to produce the thermal decoupling observed in the absence of a mixing chamber shield is 0.2 nW \cite{Schmoranzer19a}. One possibility for the origin of this heat is leakage of thermal radiation. In the absence of a mixing chamber shield, we did not observe a change in the temperature dependence of the DPO resonance frequency after the still temperature was changed from 630 to 900 mK. Thus any stray photons responsible for heating the DPO must have been radiated from surfaces warmer than the still. Indeed, when the pulse tube refrigerator was turned off, allowing the 4 K and 50 K plates to warm while the MXC and still plate temperatures were held constant, the resonance frequency of DPO8 without a hermetic shield increased (see Fig. 3 of \cite{Schmoranzer19a}). The stray photons responsible for heating must then have entered the still shield through small gaps that remained despite the use of copper tape to cover any holes. In contrast, when the pulse tube was turned off with DPO8 in the hermetic cell at 10 mK, no change in the resonance frequency was observed.

Another possibility for the origin of the heating is thermal transport by residual helium gas. However, we observe no evidence of absorption of helium on the DPO, which would cause a resonance frequency shift due to mass loading. (The effect of mass loading would dominate the elastic effect of a uniform helium layer because the shear modulus of solid helium is over 1000 times smaller than that of Si \cite{Fefferman14}, while the density of solid helium is only about 10 times smaller than that of Si.) A monolayer of helium would cause a relative frequency shift of $5 \times 10^{-8}$ \cite{Rosner03}, which would be easily observed (Fig. \ref{fig:3}). Thus heating by helium gas is less likely than heating by stray photons.

Superconducting qubits are sensitive not only to stray photons propagating in free space but also to an additional thermal decoupling mechanism: photons transmitted to the sample by microwave wiring. Very recently, Lane \emph{et al.} constructed a hermetic 3D microwave cavity containing a transmon qubit \cite{Lane19}. The relaxation and dephasing times of the qubit were measured with and without superfluid helium filling the microwave cavity. A photon bath temperature inside the cavity of 129 mK with superfluid and 150 mK without superfluid was inferred from the dephasing rate of the qubit. This degree of decoupling between the photon bath temperature and mixing chamber temperature is routinely observed in circuit QED experiments and is due to incomplete thermalization of microwave attenuators in the transmission lines leading to the cavity \cite{Yeh17}. In the present work, we placed an ECCOSORB low pass filter at the 4 K stage and used superconducting coax with no additional filters between 4 K and the mixing chamber. Evidently, electronic noise transmitted by the wiring did not significantly limit the thermalization of the DPO, as demonstrated by the excellent thermalization of our sample when inside our hermetic cell. Achieving good thermalization of attenuators and circulators at mK temperature remains a challenge in the field of quantum electronics.

In conclusion, we have demonstrated much better sample thermalization with a hermetic cell than with a simpler shield that was not leak tight. The hermetic cell allows us to cool the sample to temperatures near 10 mK. The design incorporates a simple pinch seal instead of requiring the installation of an additional pumping line in the cryostat. A shield of this design would be effective in the protection of many types of devices from heating by stray photons or residual helium, which is especially significant in dry dilution refrigerators.

We thank Gilles Pont for valuable technical advice. We acknowledge support from the ERC StG grant UNIGLASS No. 714692 and ERC CoG grant ULT-NEMS No. 647917 and from the US Office of Naval Research. The research leading to these results has received funding from the European Union’s Horizon 2020 Research and Innovation Programme, under Grant Agreement no 824109, the European Microkelvin Platform (EMP).

\end{document}